\documentstyle[psfig]{mn}

\newif\ifAMStwofonts

\newcommand{\beq}{\begin{equation}}
\newcommand{\eeq}{\end{equation}}
\newcommand{\beqn}{\begin{eqnarray}}
\newcommand{\eeqn}{\end{eqnarray}}
\def\bI{\hbox{$\,I\!\!\!\!-$}}
\def\agt{\mathrel{\raise.3ex\hbox{$>$}\mkern-14mu\lower0.6ex\hbox{$\sim$}}}
\def\alt{\mathrel{\raise.3ex\hbox{$<$}\mkern-14mu\lower0.6ex\hbox{$\sim$}}}

\ifoldfss
  \ifCUPmtlplainloaded \else
    \NewTextAlphabet{textbfit} {cmbxti10} {}
    \NewTextAlphabet{textbfss} {cmssbx10} {}
    \NewMathAlphabet{mathbfit} {cmbxti10} {} 
    \NewMathAlphabet{mathbfss} {cmssbx10} {} 
  \fi
  \ifAMStwofonts
    \ifCUPmtlplainloaded \else
      \NewSymbolFont{upmath} {eurm10}
      \NewSymbolFont{AMSa} {msam10}
      \NewMathSymbol{\upi}     {0}{upmath}{19}
      \NewMathSymbol{\umu}     {0}{upmath}{16}
      \NewMathSymbol{\upartial}{0}{upmath}{40}
      \NewMathSymbol{\leqslant}{3}{AMSa}{36}
      \NewMathSymbol{\geqslant}{3}{AMSa}{3E}

      \let\leq=\leqslant 
       
    \fi
  \fi
\fi 

\ifnfssone
  \newmathalphabet{\mathit}
  \addtoversion{normal}{\mathit}{cmr}{m}{it}
  \addtoversion{bold}{\mathit}{cmr}{bx}{it}
  \newmathalphabet{\mathbfit} 
  \addtoversion{normal}{\mathbfit}{cmr}{bx}{it}
  \addtoversion{bold}{\mathbfit}{cmr}{bx}{it}
  \newmathalphabet{\mathbfss} 
  \addtoversion{normal}{\mathbfss}{cmss}{bx}{n}
  \addtoversion{bold}{\mathbfss}{cmss}{bx}{n}
  \ifAMStwofonts
    \ifCUPmtlplainloaded \else
      \UseAMStwoboldmath
      \makeatletter
      \new@mathgroup\upmath@group
      \define@mathgroup\mv@normal\upmath@group{eur}{m}{n}
      \define@mathgroup\mv@bold\upmath@group{eur}{b}{n}
      \edef\UPM{\hexnumber\upmath@group}
      \new@mathgroup\amsa@group
      \define@mathgroup\mv@normal\amsa@group{msa}{m}{n}
      \define@mathgroup\mv@bold\amsa@group{msa}{m}{n}
      \edef\AMSa{\hexnumber\amsa@group}
      \makeatother
      \mathchardef\upi="0\UPM19
      \mathchardef\umu="0\UPM16
      \mathchardef\upartial="0\UPM40
      \mathchardef\leqslant="3\AMSa36
      \mathchardef\geqslant="3\AMSa3E

      \let\leq=\leqslant 

    \fi
  \fi
\fi 

\ifnfsstwo
  \DeclareMathAlphabet{\mathbfit}{OT1}{cmr}{bx}{it}
  \SetMathAlphabet\mathbfit{bold}{OT1}{cmr}{bx}{it}
  \DeclareMathAlphabet{\mathbfss}{OT1}{cmss}{bx}{n}
  \SetMathAlphabet\mathbfss{bold}{OT1}{cmss}{bx}{n}
  \ifAMStwofonts
    \ifCUPmtlplainloaded \else
      \DeclareSymbolFont{UPM}{U}{eur}{m}{n}
      \SetSymbolFont{UPM}{bold}{U}{eur}{b}{n}
      \DeclareSymbolFont{AMSa}{U}{msa}{m}{n}
      \DeclareMathSymbol{\upi}{0}{UPM}{"19}
      \DeclareMathSymbol{\umu}{0}{UPM}{"16}
      \DeclareMathSymbol{\upartial}{0}{UPM}{"40}
      \DeclareMathSymbol{\leqslant}{3}{AMSa}{"36}
      \DeclareMathSymbol{\geqslant}{3}{AMSa}{"3E}

      \let\leq=\leqslant 

    \fi
  \fi
\fi 

\ifCUPmtlplainloaded \else
  \ifAMStwofonts \else 
    \def\upi{\pi}
    \def\umu{\mu}
    \def\upartial{\partial}
  \fi
\fi

\title[Dynamical Bar-mode Instability]{Dynamical bar-mode instability of differentially rotating stars: 
Effects of equations of state and velocity profiles}
\author[Shibata et al.]
       {Masaru Shibata, Shigeyuki Karino, and Yoshiharu Eriguchi \\ 
	Department of Earth Science and Astronomy, 	
	Graduate School of Arts and Sciences,~University of Tokyo,\\
	Komaba, Meguro, Tokyo 153-8902, Japan}
\date{Accepted ???? Month ??,
      Received ???? Month ??;
      in original form 2002 April 1}

\pagerange{\pageref{firstpage}--\pageref{lastpage}}
\pubyear{????}

\begin{document}

\maketitle

\label{firstpage}

\begin{abstract}

As an extension of our previous work,
we investigate the dynamical instability 
against nonaxisymmetric bar-mode deformations of differentially rotating 
stars in Newtonian gravity varying the equations of state and
velocity profiles. We performed 
the numerical simulation and the followup linear stability analysis 
adopting polytropic equations of state with the polytropic
indices $n=1$, 3/2, and 5/2 and with 
two types of angular velocity profiles (the so-called $j$-constant-like and 
Kepler-like laws). 
It is confirmed that rotating stars of a high degree of 
differential rotation are dynamically unstable against the bar-mode 
deformation, even for the ratio of the kinetic energy to the gravitational 
potential energy $\beta$ of order 0.01. The criterion for onset of 
the bar-mode dynamical 
instability depends weakly on the polytropic index $n$ and the angular 
velocity profile as long as the degree of differential rotation is high. 
Gravitational waves from the final nonaxisymmetric quasi-stationary states are 
calculated in the quadrupole formula.
For proto-neutron stars of mass 
$1.4M_{\odot}$, radius $\sim 30$ km and $\beta \alt 0.1$, 
such gravitational waves have the frequency of $\sim$ 600--1,400 Hz, 
and the effective amplitude is larger than $10^{-22}$ at a distance of 
about $100$ Mpc irrespective of $n$ and the angular velocity profile. 
\end{abstract}


\section{INTRODUCTION}

In our previous paper, we studied dynamical bar-mode stabilities 
of differentially rotating stars in Newtonian gravity \cite{SKE}.
In that study, we adopted 
a polytropic equation of state with the polytropic index $n=1$ and the 
so-called ``$j$-constant-like'' angular velocity profile in which 
the magnitude of the angular velocity decreases as $\varpi^{-2}$ at large 
values of $\varpi$, where $\varpi$ denotes the cylindrical radius. 
We found that rotating stars of a high degree of
differential rotation are dynamically unstable even 
for $\beta \equiv |T/W| \sim 0.03$, where $T$ and $W$ are rotational and 
gravitational potential energies. This value is much smaller than the 
long-believed criterion of $\beta \approx 0.27$ for onset of the
bar-mode dynamical instability of rotating stars (see Shibata et al. 2002 
for a review). We also found that after the instability
sets in, such unstable rotating stars with 
$0.03 \alt \beta \alt 0.15$ eventually
settle down to nonaxisymmetric ellipsoidal quasi-stationary states. 

However, there are two questions which have not been
answered in the previous work \cite{SKE}.
One is associated with our choice of the $j$-constant-like angular 
velocity profile. It is well known that accretion disks around a central 
body with constant specific angular momentum are unstable against the 
Papaloizou-Pringle instability \cite{PP}. On the other hand, the accretion 
disks are stable if the velocity profile is Kepler-like.
One could claim that the bar-mode instability which we found 
would not be universal and might set in only for 
the very special rotational profile such as the $j$-constant-like law
as in the Papaloizou-Pringle instability. In addition,
we focused only on a stiff equation of state with 
$n=1$ in the previous paper,
so that one could also ask if the instability sets in for softer 
equations of state. 

To answer these questions, we have performed numerical simulations
of differentially rotating stars 
varying the polytropic index and the angular velocity profile.
In this paper, we report the numerical results.
We will show that irrespective of 
the polytropic index $n$ and the angular velocity profile,
a rotating star of 
a high degree of differential rotation is dynamically 
unstable against the bar-mode deformation even 
for the case that $\beta$ is of order 0.01. We will
also show that an unstable star of a small value of $\beta$
eventually settles down to 
a nonaxisymmetric quasi-stationary state, which 
is a strong emitter of quasi-periodic gravitational waves. 

The paper is organized as follows. In Section 2, we describe our methods in 
numerical analysis. In Section 3, the
numerical results are presented.
Section 4 is devoted to a summary and discussion. 
Throughout this paper, we use the geometrical
units of $G=c=1$ where $G$ and $c$ denote 
the gravitational constant and the light velocity. 

\section{METHOD}

\subsection{Differentially rotating axisymmetric stars} 

We set a differentially rotating star in equilibrium, and
investigate the dynamical stability.
Rotating stars in equilibrium are modeled using the polytropic 
equations of state as $P=K\rho^{\Gamma}$ where $P$, $\rho$, $K$ and 
$\Gamma=1+1/n$ denote the pressure, density, polytropic constant and 
adiabatic index. In this paper,
we choose $n=1$, 3/2, and 5/2 ($\Gamma=2$, 5/3, and 7/5).

As the angular velocity profile $\Omega(\varpi)$, 
we choose the so-called $j$-constant-like law as 
\beq
\Omega = {\Omega_0 A^2 \over \varpi^2 + A^2},\label{omegaj}
\eeq
and the so-called Kepler-like law as 
\beq
\Omega = \Omega_0 \biggl[{A^2 \over \varpi^2 + A^2}\biggr]^{3/4},
\label{omegak}
\eeq
where $A$ is a constant, and $\Omega_0$ the angular velocity at the symmetric 
axis. The parameter $A$ controls the steepness of the angular velocity profile:
For smaller values of $A$,
the profile is steeper and for $A \rightarrow \infty$, the 
rigid rotation is recovered. In the present work, the values of $A$ are
chosen among the range 
$0.1 \leq \hat A \equiv A/R_{\rm eq} \leq 1$ where $R_{\rm eq}$ is the 
equatorial radius of rotating stars. For the rotation laws 
(\ref{omegaj}) and (\ref{omegak}), $\Omega$ at large cylindrical radius
asymptotically 
behaves as $\varpi^{-2}$ and $\varpi^{-3/2}$. This is the reason why we refer 
to the profiles (\ref{omegaj}) and (\ref{omegak}) as
the $j$-constant-like and Kepler-like laws. 

In the limit of $A \rightarrow 0$ with the profile (\ref{omegaj}), 
the specific angular momentum becomes constant everywhere and $\Omega$ 
diverges at $\varpi=0$. We note that this profile has been often used 
in studies of nonaxisymmetric instabilities in tori and annuli 
\cite{PP,GN,TH,AT}. In this paper, however, we do not consider tori and annuli 
and focus only on spheroidal
stars for which the density is not zero at $\varpi=0$. 
Thus, we cannot adopt this limiting profile.  

In terms of $\beta=T/|W|$ and $\hat A$, one rotating star is determined
for a given rotational profile and polytropic index. 
Thus, in the following, we often
refer to these two parameters to specify a rotating 
star. Here, $T$ and $W$ are defined as 
\beqn
&&T={1 \over 2}\int d^3x \rho \varpi^2 \Omega^2,\\
&&W={1 \over 2}\int d^3x \rho \phi,
\eeqn
where $\rho$ and $\phi$ are the mass density and the Newtonian gravitational 
potential. To specify a particular model, 
we may choose the axis ratio of the rotating 
stars $C_a$ instead of $\beta$.
Here $C_a$ is defined as the ratio of the polar radius $R_p$ to the 
equatorial radius $R_{\rm eq}$, i.e., $C_a=R_p/R_{\rm eq}$. For the equations 
of state and the angular velocity profiles that we study in this paper, 
the value of $C_a$ monotonically decreases with increase of $\beta$
for a given set of $\hat A$ and $n$. This is the reason that
$C_a$ can be a substitute for $\beta$. 

\subsection{Dynamical stability investigation}

To investigate the dynamical stability against nonaxisymmetric 
bar-mode deformations, 
we have performed the numerical simulation as well as the 
linear stability analysis. Below
we explain the methods for our numerical computation separately.

\subsubsection{Numerical simulation}

In the hydrodynamic simulation, we initially superimpose a nonaxisymmetric
density perturbation to an axisymmetric equilibrium star.  
We focus mainly on a fundamental bar-mode, and simply add the 
node-less density perturbation of the form 
\beq
\delta \rho = \delta \cdot \rho_0(\varpi,z) {x^2 - y^2 \over R_{\rm eq}^2},
\label{pert}
\eeq
where $\rho_0(\varpi,z)$ denotes the density of the axisymmetric configuration 
and $\delta$ constant. Throughout this paper, we choose $\delta=0.1$. 
For simplicity, the velocity is left to be unperturbed 
at $t=0$. The growth of the bar-mode can be followed by monitoring the 
distortion parameter as 
\beqn
\eta \equiv ( \eta_{+}^2 + \eta_{\times}^2)^{1/2}, 
\eeqn
where 
\beqn
&&\eta_+ \equiv {I_{xx} - I_{yy} \over I_{xx} + I_{yy}},\\
&&\eta_{\times} \equiv {2I_{xy} \over I_{xx} + I_{yy}},
\eeqn
and $I_{ij}~(i,j=x,y,z)$ denotes the quadrupole moment defined by 
\beqn
I_{ij} = \int d^3x \rho x^i x^j. 
\eeqn
Here, $x^i=(x, y, z)$. Simulations are performed using a
3D numerical hydrodynamic implementation 
in Newtonian gravity \cite{SON} (see also Shibata 2000 for results of
various test simulations with the identical
hydrodynamic numerical scheme but in general relativity).
We adopt a fixed uniform grid with 
size $141\times 141\times 141$ in $x-y-z$, which covers an equatorial radius 
by 50 grid points initially. We also performed test simulations with size 
$71\times 71\times 71$ (i.e., the grid spacing becomes twice larger)
for several selected cases and confirmed that the results depend weakly 
on the grid resolution. We assume a reflection symmetry with respect to the 
equatorial plane. Since several rotating stars
that we picked up have a flattened 
configuration, we set the grid spacing of $z$ half of that of $x$ and $y$. 

\subsubsection{Linear stability analysis}

In the linear stability analysis, we employ the scheme developed by
Karino et al. (2000, 2001).  The Euler perturbations of the physical 
quantities are replaced by functions of the form 
$f(r, \theta)e^{im\varphi-i\omega t}$ in the the linearized 
hydrodynamic equations. Here $m$ is an azimuthal mode number. As a 
result, the problem reduces to the eigenvalue problem for an eigenvalue 
$\omega$ and the corresponding eigenfunctions of the perturbed quantities. 
We assume an adiabatic relation between the Euler perturbation of the pressure 
and that of the density. Since the fundamental mode of the $m = 2$ or bar-type 
oscillations is node-less, we have checked whether the obtained eigenfunctions 
satisfy that condition. In this paper, we have analyzed the stability of the 
equilibrium configurations of $n=1$ polytropes for two rotation 
laws (\ref{omegaj}) and (\ref{omegak}) with several values of $\hat A$. 

\section{NUMERICAL RESULTS}

\subsection{Dynamical Bar-mode Stability}

\subsubsection{Results by numerical simulation}

The dynamical stability is
studied for various combination of $\Gamma$, $\hat A$ 
and $\beta$, and for two angular velocity profiles
in the numerical simulation. In Figures 1 and 2, we 
summarize the results with regard to the dynamical bar-mode stability 
for the $j$-constant-like and Kepler-like angular velocity profiles, 
respectively. Here, the circles (crosses) denote that the stars of a
given set of $\hat A$ and $\beta$ are dynamically unstable (stable).
We focus only on the spheroidal stars which are located
below the dashed curves plotted in Figures 1 and 2.
(If the value of $\beta$ is larger than that on
this curve for a given value of $\hat A$, the star is toroidal.)

\begin{figure}
\vspace*{-6mm}
\begin{center}
\leavevmode
\psfig{file=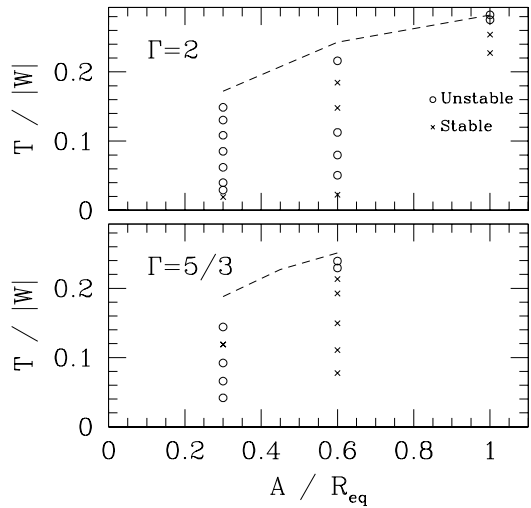,width=3.2in,angle=0}	
\end{center}
\vspace*{-8mm}
\caption{The dynamical bar-mode stability is shown in the plane of 
$\beta$ and $\hat A$ for the $j$-constant-like angular velocity profile. 
The circles and crosses denote that the rotating stars are unstable and 
stable, respectively. The dashed curves denote the boundary
which distinguishes the spheroidal stars from the toroidal stars
(spheroidal stars are located below the curves).
}
\end{figure}

\begin{figure}
\vspace*{-6mm}
\begin{center}
\leavevmode
\psfig{file=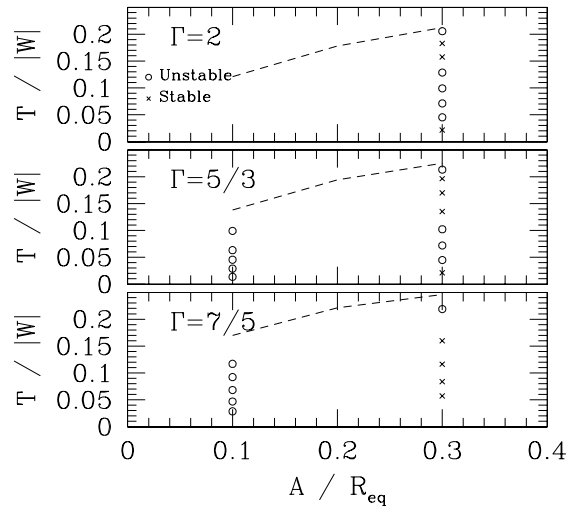,width=3.2in,angle=0}
\end{center}
\vspace*{-8mm}
\caption{The same as Figure 1 but for the Kepler-like angular 
velocity profile. 
}
\end{figure}

It is found that many rotating stars of a high degree of differential rotation 
as $\hat A =0.1$ and 0.3 are dynamically unstable even for $\beta$ of 
order 0.01. In the case of the $j$-constant-like angular velocity profile, 
most of rotating stars with $\beta \agt 0.01$ and $\hat A=0.3$
are dynamical unstable 
both for $\Gamma=2$ and 5/3. In the case of the Kepler-like 
angular velocity profile, the threshold of $\hat A$ for onset of
the instability for a given set
of $\beta$ and $\Gamma$ appears to be slightly 
smaller than that in the $j$-constant-like case. This is  
because the Kepler-like angular velocity profile is not as steep as the
$j$-constant-like one for the same value of $\hat A$.
Indeed, for $\Gamma=2$ and 5/3 and for $\hat A=0.3$,
there is a wide parameter space around
$\beta \sim 0.15$ in which the stars are stable
against the bar-mode deformation in the Kepler-like case. 
Such a wide parameter space is absent in the $j$-constant-like
angular velocity profile for $\hat A=0.3$. 
However, for $\hat A = 0.1$, the stars in a wide range of $\beta$ 
are unstable even in the Kepler-like angular 
velocity profile. Thus, {\it rotating stars of a high degree of
differential rotation with $\beta$ of order 0.01 
are dynamically unstable 
even in the Kepler-like angular velocity profile.} 

The threshold of $\hat A$ for onset of the dynamical instability 
for a given angular velocity profile 
also appears to depend on $\Gamma$: For a smaller value of $\Gamma$, 
the threshold of $\hat A$ is smaller. For example,
compare the results with $\Gamma=5/3$ and 7/5 in Fig. 2.
For $\Gamma=5/3$, the stars of $\hat A=0.3$ and
$0.03 \alt \beta \alt 0.1$ are unstable, but for $\Gamma=7/5$,
all the stars with $\beta < 0.2$ that we studied are stable for $\hat A =0.3$. 
The reason is that the rotating stars of smaller values of
$\Gamma$ have a more centrally condensed structure and as a result 
the effective degree of differential rotation of the central core
for the stars of soft equations of state can become very high only
for a sufficiently small value of $\hat A$. 

The rotating stars of a high degree of differential rotation
are also dynamically unstable for a high value of $\beta~(\agt 0.2)$
(see the results of $\hat A = 0.6$ and 1 for the $j$-constant-like law and
of $\hat A=0.3$ for the Kepler-like law). 
The axial ratio $C_a$ of these stars is 
very small ($C_a \alt 0.3$). In particular, for the unstable
stars of the Kepler-like angular velocity profile with $\beta \agt 0.2$
and $\hat A=0.3$, $C_a \sim 0.15$.
Thus, they have an almost toroidal  shape. 
On the other hand, for the unstable stars with a small value of
$\beta~(\alt 0.10)$, $C_a \agt 0.5$ and, hence, the shape is
not significantly toroidal. 


An interesting feature is found for $\Gamma=2$ and $\hat A=0.6$ of the
$j$-constant-like angular velocity profile, and for $\Gamma=5/3, 2$ and 
$\hat A=0.3$ of the Kepler-like angular velocity profile. 
In these cases, the stability does not change monotonically with increase
of $\beta$: 
(1) stars of a sufficiently small value of $\beta~(\alt 0.01)$ are stable
(we have not carried out simulations for $\beta < 0.01$, but
since the value of $C_a$ with $\beta \alt 0.01$ is larger than 0.9 
(i.e., the star is almost spherical), we assume that the stars with
$\beta \alt 0.01$ are stable), 
(2) stars of $0.01 \alt \beta \alt 0.1$ are unstable, 
(3) stars of $0.1 \alt \beta \alt 0.2$ are stable
again, and 
(4) stars of a sufficiently large value of $\beta~(\agt 0.2)$
are unstable again. 

It has been widely believed that the value of $\beta$ is a good 
indicator to distinguish unstable stars from stable ones. However, the 
examples shown here illustrate 
that $\beta$ is not a good indicator for determination of the dynamical
stability of rotating stars of a high degree of differential rotation.

The reason for the fact
that the stability does not change monotonically with $\beta$
is not clear. 
The likely reason is that the mode associated with the
stability for the high value of $\beta$ is 
different from that for the small value. 
However, this interpretation cannot be proved at present 
because of the following reason.
The property of the unstable modes for the high value of $\beta$ 
such as the perturbed density profile and the magnitude itself
($\beta \agt 0.2$) is essentially the same as that of the well-known 
bar-mode (i.e, the $m=2$ toroidal mode, see Chandrasekhar 1969).
Hence, we identify the unstable modes of high value of $\beta$
as the $f$ mode. 
On the other hand, for the small value of $\beta$, the perturbed 
density profile is also node-less, and furthermore, 
the real part of the eigen frequency is approximately
proportional to $(M/R_{\rm eq}^3)^{1/2}$
(see Figures 10 and 11). These are the properties that 
the $f$ mode should have. 
This fact suggests that the unstable modes for the
small value of $\beta$ might be also the $f$ mode. 

We suspect that there is something different between two unstable 
modes of high and low values of $\beta$. 
However, to find the difference, 
it is necessary to precisely understand the definition of the 
$f$ mode against the bar-deformation for {\it differentially rotating stars}.
At present, it is not clear for us what properties, 
besides the node-less density profile and 
the eigen frequency of order of magnitude $\sim (M/R_{\rm eq}^3)^{1/2}$, 
characterize the $f$ mode. Namely, we do not know what to prove.  
Precisely defining the $f$ mode for differentially rotating stars 
is beyond scope of this paper, and is left as a problem in future. 

Before closing this section, we address the following point. 
As pointed out by Centrella et al. (2001),
the dynamical instability for $m=1$ mode may set in for
rotating stars of a high degree of differential rotation.
In this paper, we have not found the evidence that
such instability sets in. This seems to be due to
the fact that we adopted only the stiff equations of state.
Centrella et al. (2001) studied the instability for
differentially rotating stars of a very
soft equation of state with $n=10/3$. In such a soft equation of state,
the instability associated with the $m=1$ mode plays an important
role. However, this is not likely the case in the stiff equations of
state. Indeed, recently, Saijo et al. (2003) have pointed out 
that the instability of the $m=1$ mode sets in only for 
differentially rotating stars of very soft equations 
of state ($n \agt 2.5$) and of a large value of $\beta~(\agt 0.15)$.

\subsubsection{Results by linear analysis}

In Figure 3, we display the numerical
results of the linear stability analysis for
the same equilibrium models as shown in Figures 1 and 2 for the $n=1$ 
polytrope. These figures show that the results by the numerical 
simulation agree well with those by the linear stability analysis.
This fact confirms the conclusion that rotating stars of
a high degree of 
differential rotation are dynamically unstable even for the
small value of $\beta$ 
irrespective of the angular velocity profile.
We note that the solid curve in the upper panel of Figure 3 
denotes the threshold of the dynamical stability (i.e., above this
curve the star is dynamically unstable) against
the well-known bar-mode (Chandrasekhar 1969) 
which was calculated and reported by Karino and Eriguchi (2003). 


\begin{figure}
\begin{center}
\leavevmode
\leavevmode
\psfig{file=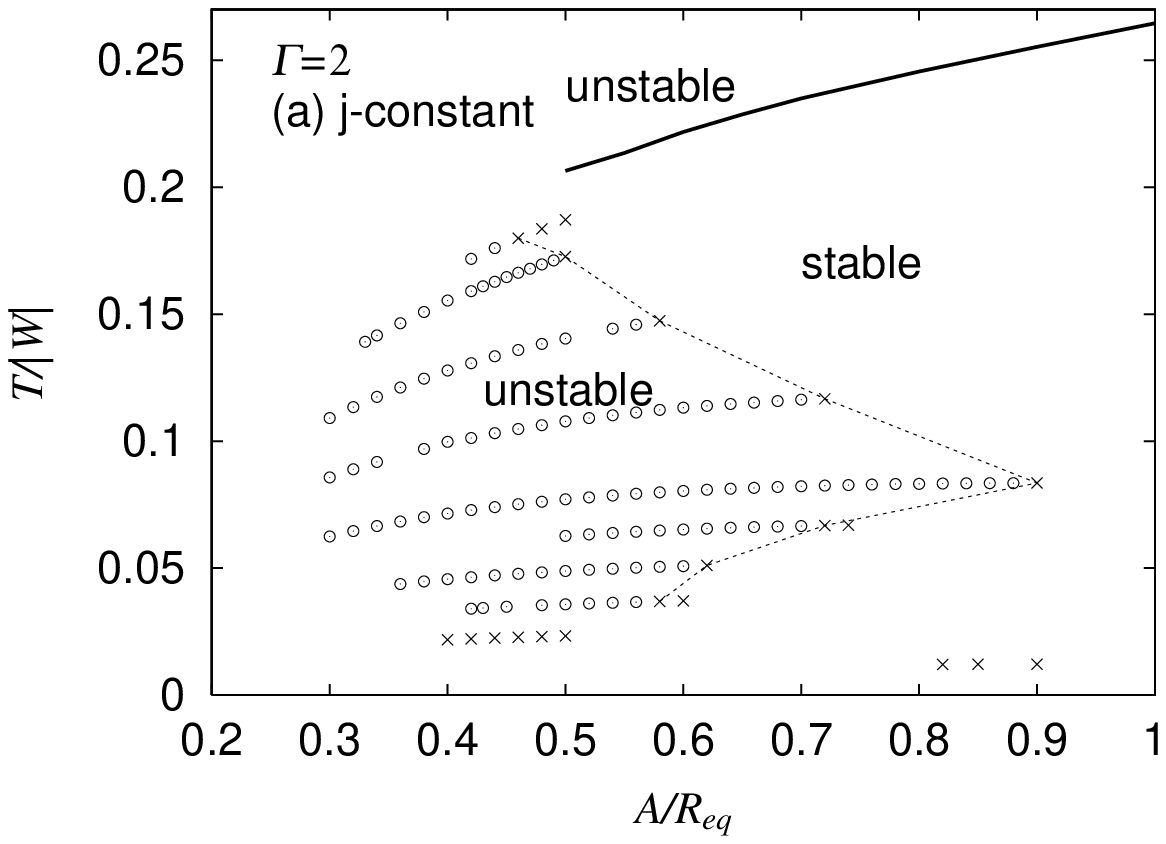,width=3.in,angle=0}
\leavevmode
\psfig{file=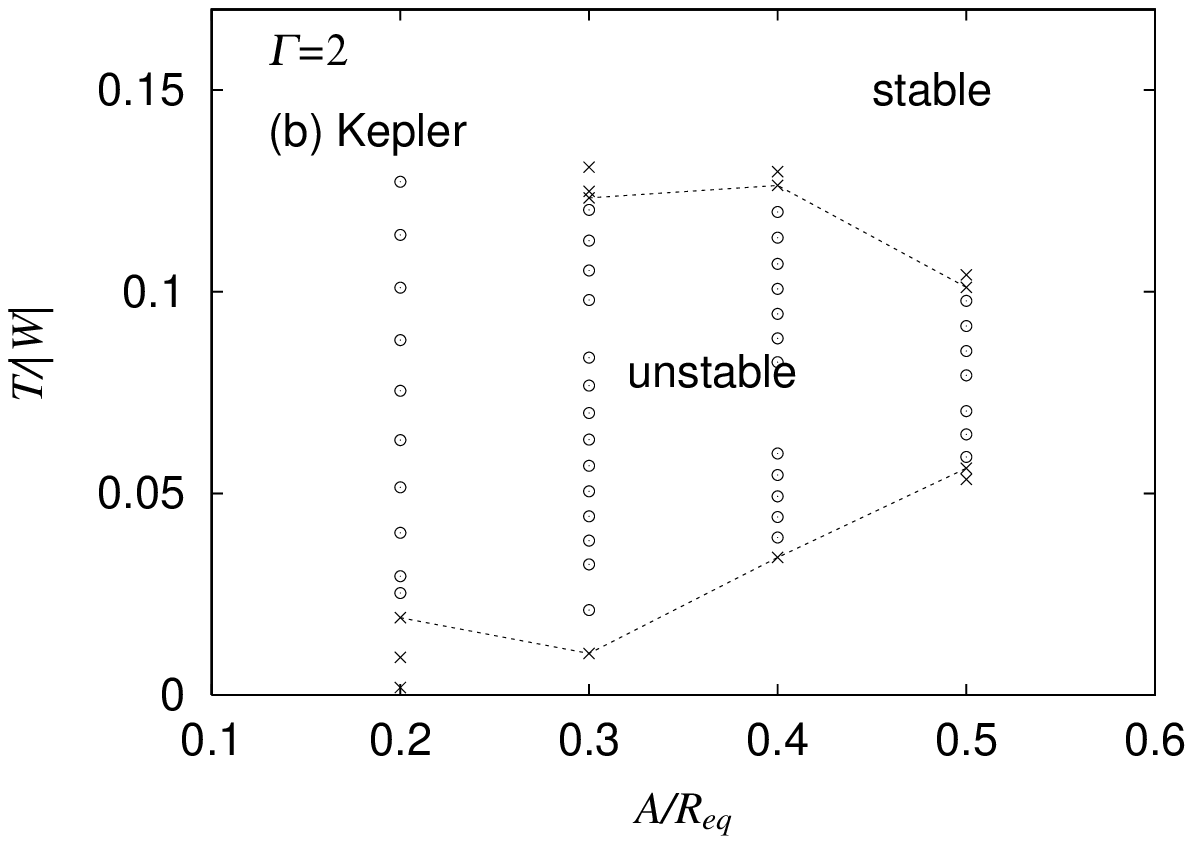,width=3.in,angle=0}
\end{center}
\caption{The results of the linear stability analysis are shown in the
plane of $\beta$ and $\hat A$ (a) for the $j$-constant-like 
and (b) for the Kepler-like angular velocity profiles.
The circles and crosses denote that the rotating stars are unstable and 
stable, respectively. The solid curve in the upper panel 
denotes the threshold of the dynamical instability against the 
well-known bar-mode which was calculated 
in Karino and Eriguchi (submitted). The dotted curves 
denote the approximate threshold of the stability for the small value of
$\beta$. 
}
\end{figure}

\subsection{Fate of unstable stars}

In this section, we focus on rotating stars of the Kepler-like 
angular velocity profile, since the results for the $j$-constant-like angular 
velocity profile with $\Gamma=2$ are shown in the previous paper \cite{SKE} 
and, moreover, we have found that
the results for $\Gamma=5/3$ show qualitatively identical features.

In Figures 4--7, we display time evolution of $\eta$ as a function of 
$\Omega_0 t$ for $(\Gamma, \hat A)=(2, 0.3)$, (5/3, 0.3), (5/3, 0.1), and 
(7/5, 0.1). As shown here, the value of
$\eta$ does not reach the value of order 1 but 
saturates at order 0.1. This implies that the growth of the instability 
saturates at a weakly nonlinear stage. 
We note that for unstable stars with $\beta \agt 0.2$ and $\hat A=0.3$,
the value of $\eta$ increases to $\sim 1$ irrespective of $\Gamma$,
implying that a highly deformed star is produced. 
There are a number of numerical works in which such a highly nonaxisymmetric 
structure is the outcome after onset of dynamical instability of
differentially rotating stars with a high value of $\beta$ 
(e.g., Williams \& Tohline 1987, 1988; Houser \& Centrella 1996).
The outcome in the simulations
for unstable stars of a high value of $\beta$ 
found in the present numerical computation is qualitatively the same as
that in previous papers. 
Thus, we do not discuss the results for such cases in the following. 

\begin{figure}
\vspace*{-6mm}
\begin{center}
\leavevmode
\psfig{file=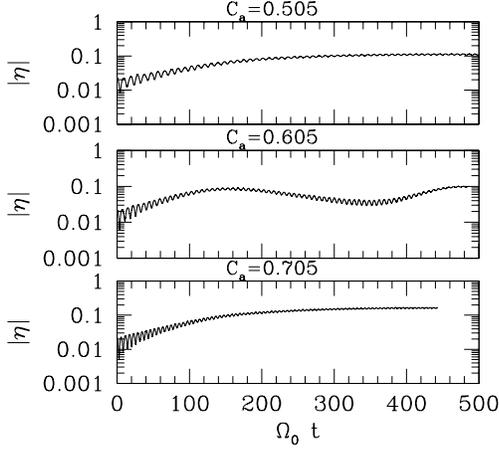,width=2.7in,angle=0}
\end{center}
\vspace*{-8mm}
\caption{Time evolution of $\eta$ as a function of $\Omega_0 t$
for $\Gamma=2$ 
and $\hat A=0.3$, and for the Kepler-like angular velocity profile. 
}
\end{figure}
\begin{figure}
\vspace*{-6mm}
\begin{center}
\leavevmode
\psfig{file=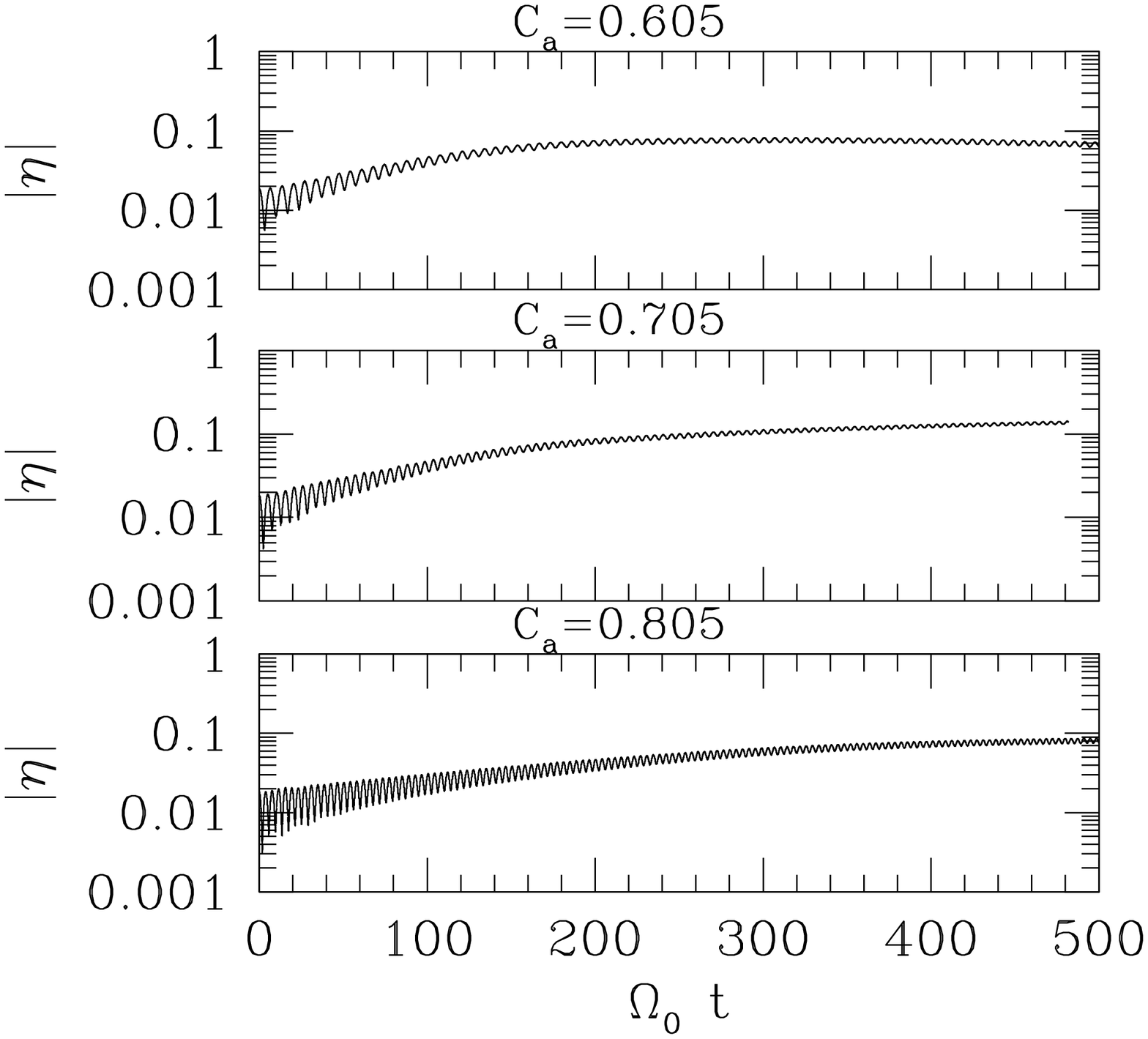,width=2.7in,angle=0}
\end{center}
\vspace*{-8mm}
\caption{The same as Figure 4 but for $\Gamma=5/3$ and $\hat A=0.3$. 
}
\end{figure}
\begin{figure}
\vspace*{-6mm}
\begin{center}
\leavevmode
\psfig{file=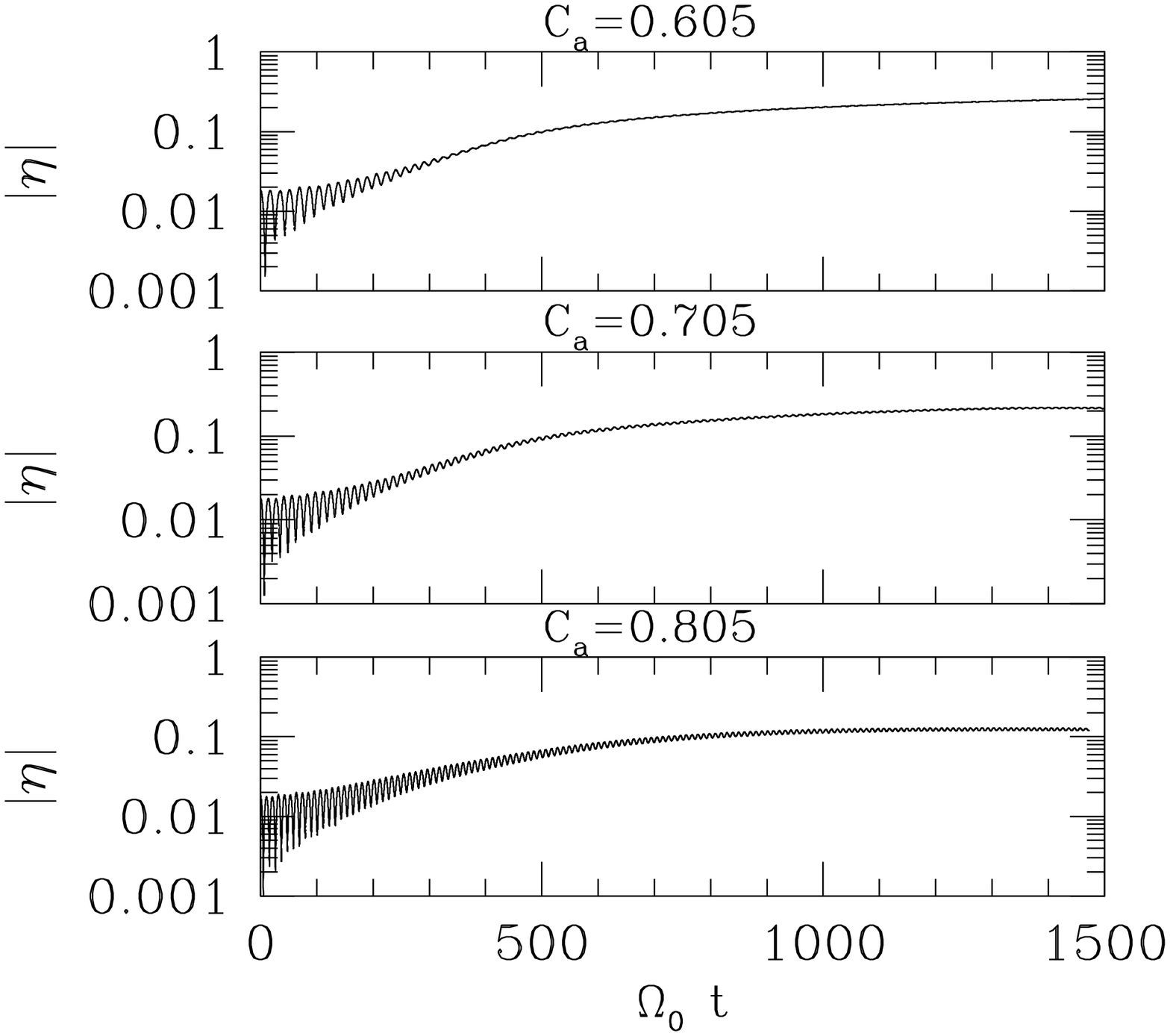,width=2.7in,angle=0}
\end{center}
\vspace*{-8mm}
\caption{The same as Figure 4 but for $\Gamma=5/3$ and $\hat A=0.1$. 
}
\end{figure}

\begin{figure}
\vspace*{-6mm}
\begin{center}
\leavevmode
\psfig{file=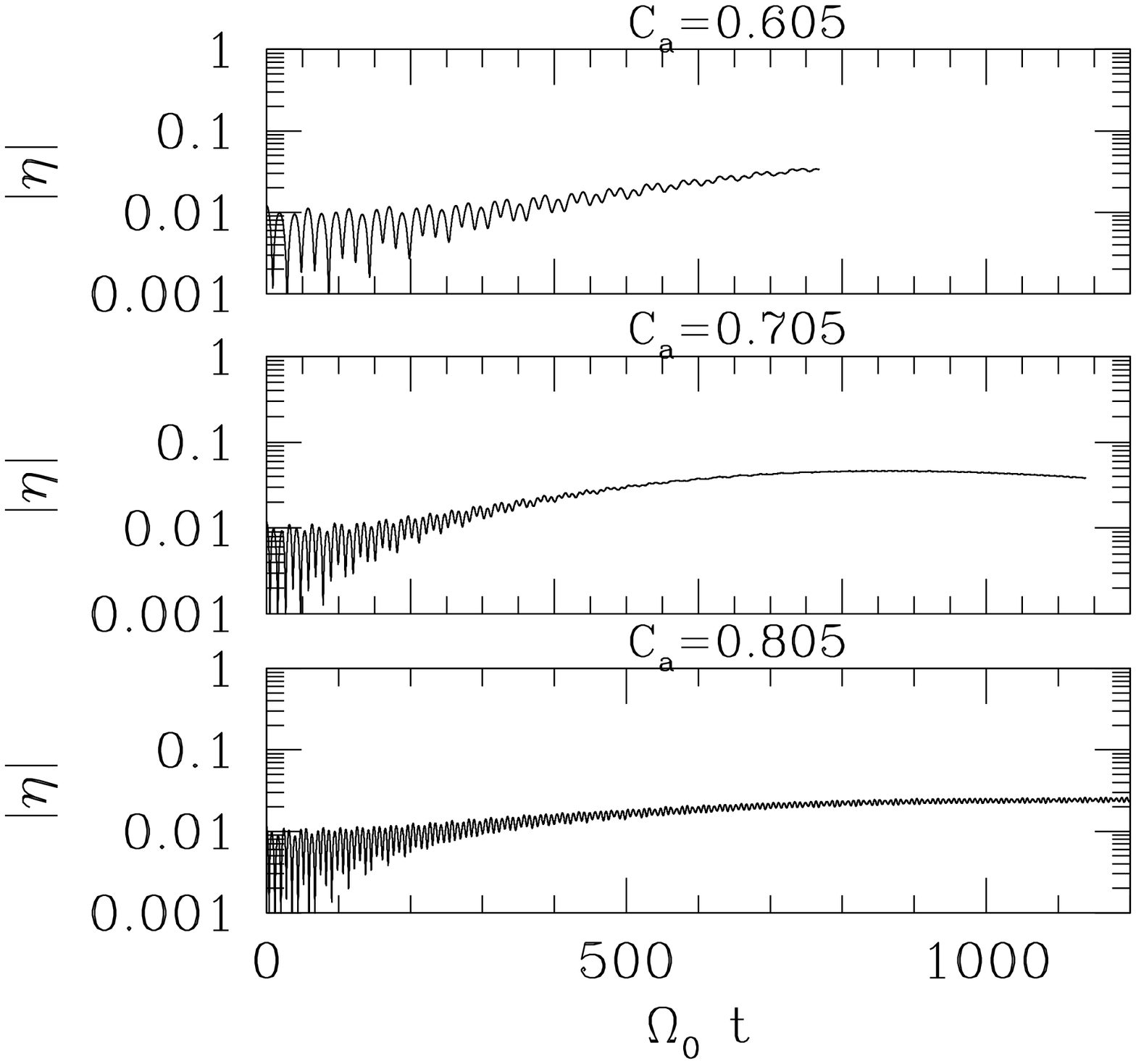,width=2.7in,angle=0}
\end{center}
\vspace*{-8mm}
\caption{The same as Figure 4 but for $\Gamma=7/5$ and $\hat A=0.1$. 
}
\end{figure}

After the perturbation saturates, the amplitude of $\eta$ settles down 
toward a value of order 0.1. The approximate final values of $\eta$ are
\beqn
\eta_{f} \left\{
\begin{array}{ll}
\sim 0.1 & {\rm for}~~(\Gamma, \hat A) = (2, 0.3), \\ 
\alt 0.1 & {\rm for}~~(\Gamma, \hat A) = (5/3, 0.3), \\ 
\sim 0.2 & {\rm for}~~(\Gamma, \hat A) = (5/3, 0.1), \\ 
\sim 0.03& {\rm for}~~(\Gamma, \hat A) = (7/5, 0.1). 
\end{array}
\right.
\eeqn
Comparing the results of 
$(\Gamma, \hat A)=(5/3, 0.3)$ and (5/3, 0.1) for the identical value
of $C_a$, it is 
found that for the smaller value of $\hat A$, $\eta_f$ is 
larger. This implies that a stronger degree of differential 
rotation makes the magnitude of the 
nonaxisymmetric deformation larger.

It is also 
found that for the smaller value of $\Gamma$, the final value of $\eta$ is 
smaller. This is because the stars of smaller values of $\Gamma$ have a more 
centrally-condensed structure and, hence, their effective steepness of the
differential rotation is smaller for smaller value of $\Gamma$ for 
a given set of $C_a$ and $\hat A$. 

In Figures 8--9, we display the snapshots of the density contour curves 
in the equatorial plane at selected time steps for 
$(\Gamma, \hat A, C_a)=(2, 0.3, 0.7)$ and (5/3, 0.3, 0.8).
The value of $\beta$ is about 0.071 and 0.045, respectively. 
In both cases, the nonaxisymmetric perturbation initially provided 
grows, changing the shape of the
rotating stars to be ellipsoidal. However, the 
perturbation does not grow to the highly nonlinear stage
and, hence, neither a 
spiral arm nor a large bar is formed in contrast to
the outcome in the simulations with a high value of $\beta \agt 0.2$.
Instead, the slightly deformed 
ellipsoid is the final outcome. Since the deformed ellipsoid is almost 
stationary, $\eta_+$ and $\eta_{\times}$
oscillate quasi-periodically in the late phase of the 
simulations. This result is qualitatively the same as for the stars of the 
$j$-constant-like angular velocity profile \cite{SKE}. 

\begin{figure}
\vspace*{-6mm}
\begin{center}
\leavevmode
\psfig{file=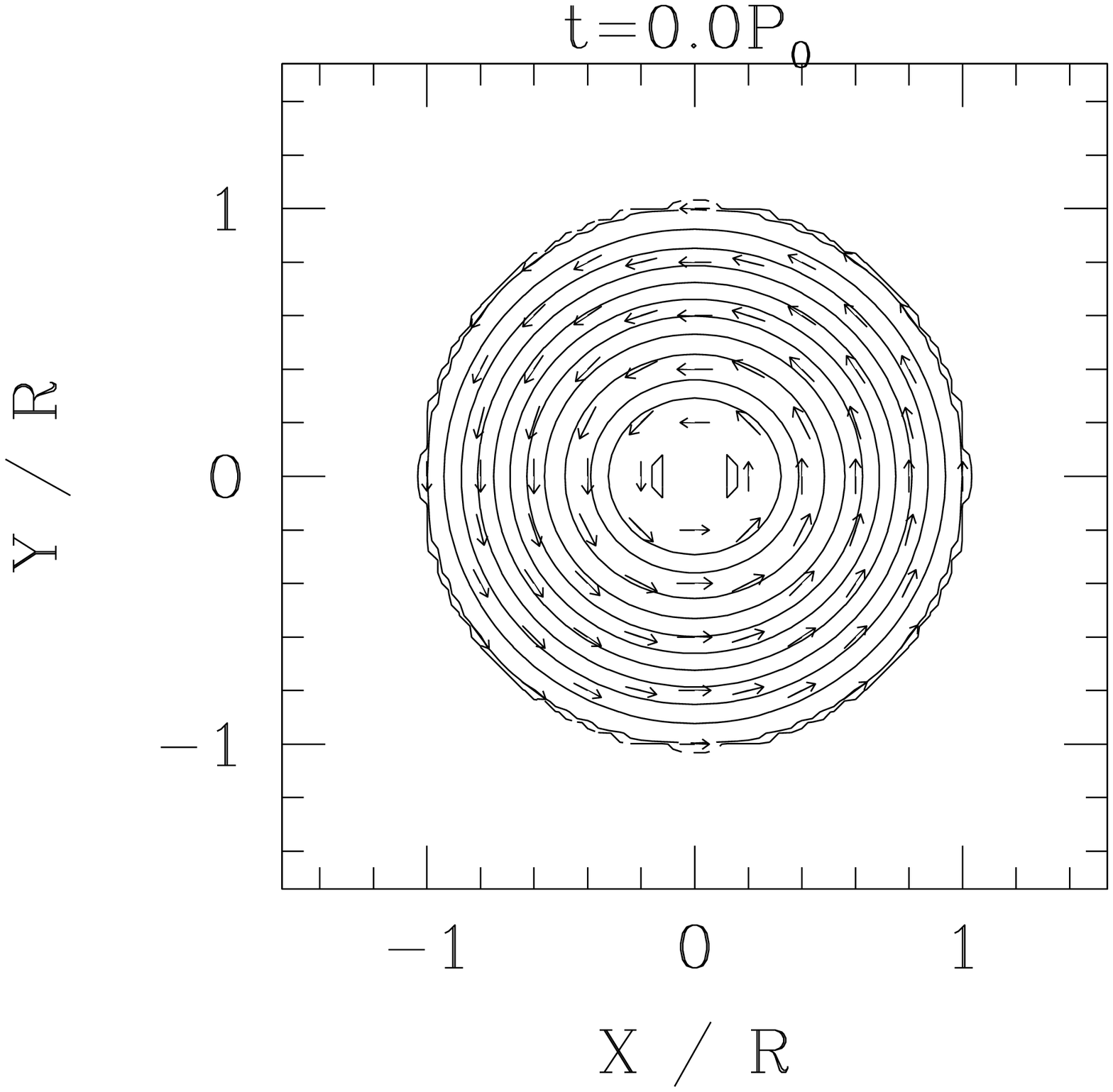,width=1.5in,angle=0}
\leavevmode
\psfig{file=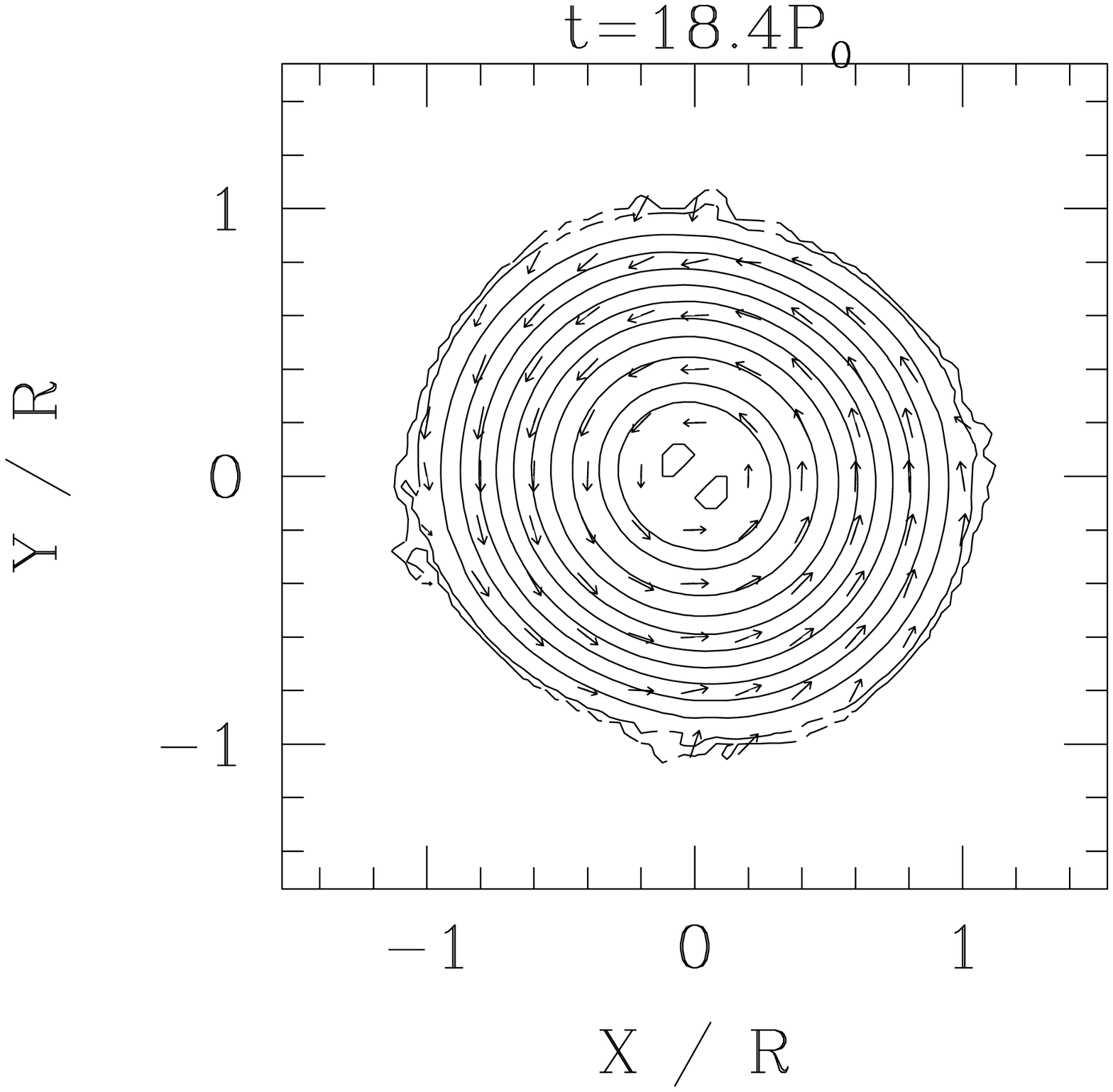,width=1.5in,angle=0}
\leavevmode
\psfig{file=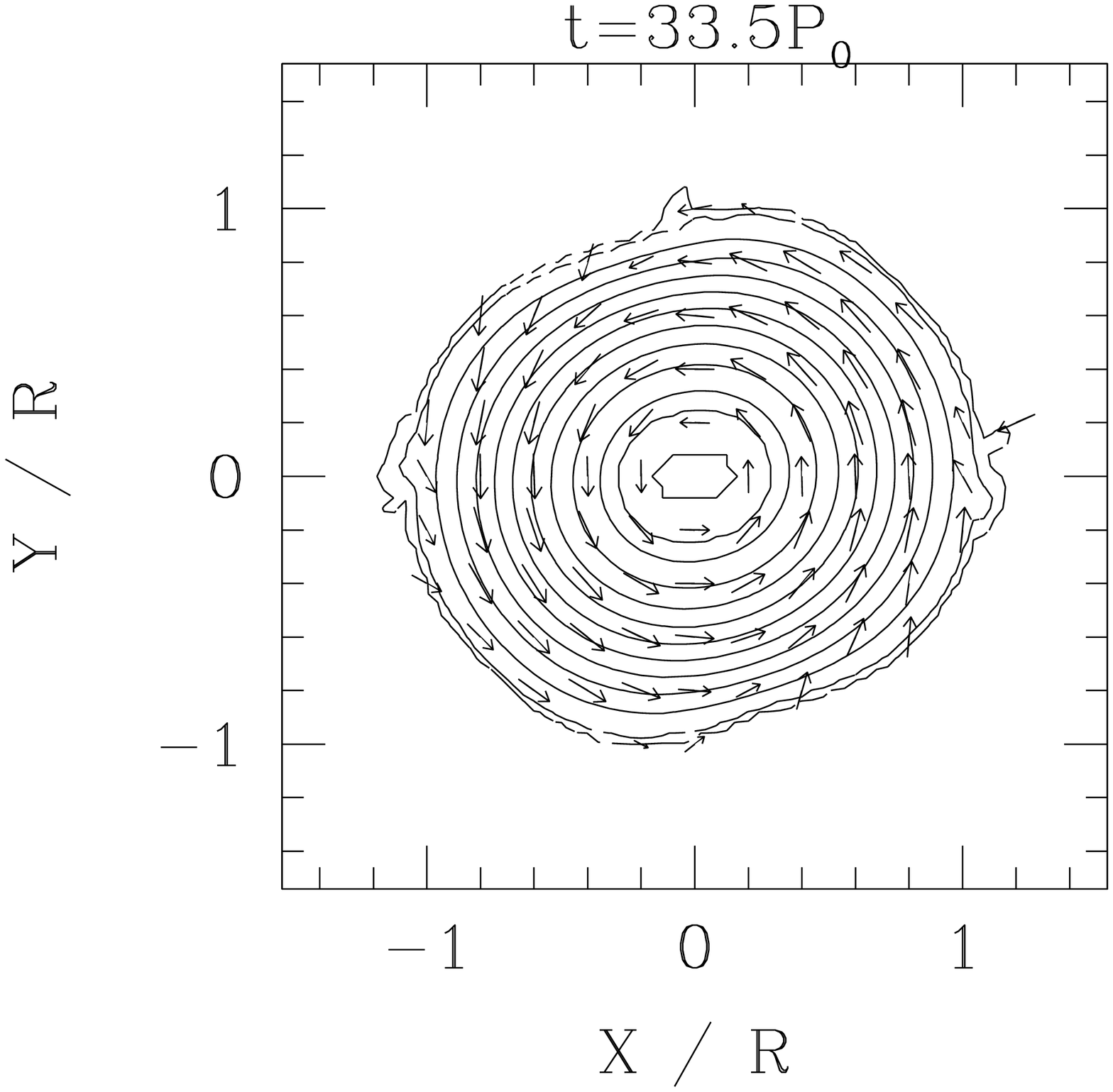,width=1.5in,angle=0}
\leavevmode
\psfig{file=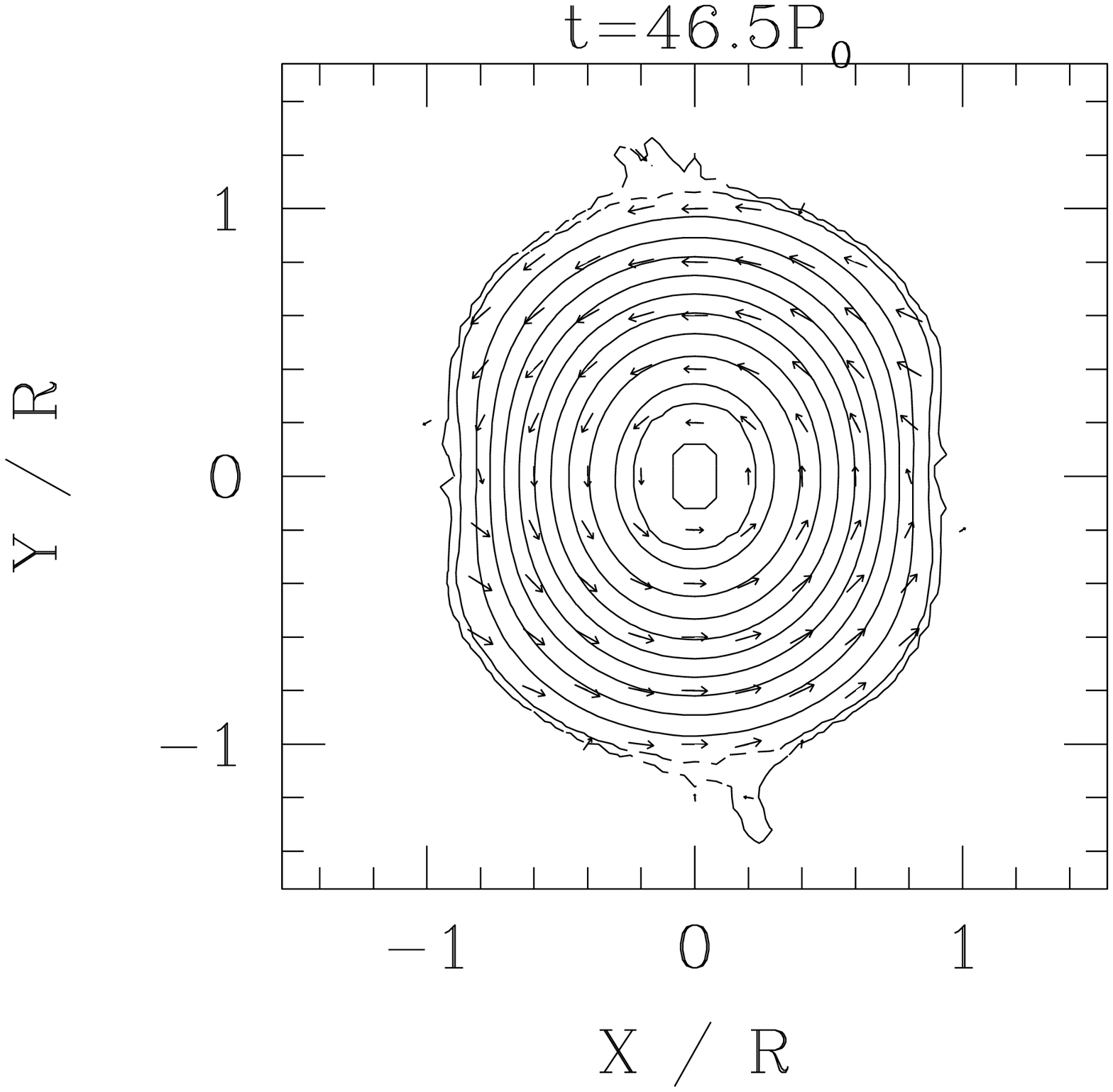,width=1.5in,angle=0}
\end{center}
\vspace*{-8mm}
\caption{The density contour curves at selected time steps 
for $\Gamma=2$, $\hat A=0.3$, and $C_a=0.7$ and 
for the Kepler-like angular velocity profile. Here, $P_0$ is 
$2\pi/\Omega_0$. The contour curves are drawn for 
$\rho/\rho_{\rm max}=0.95$, 0.9, 0.8, 0.7, 0.6, 0.5, 0.4,
0.3, 0.2, 0.1 0.01 \& 0.001,
where $\rho_{\rm max}$ denotes the maximum density at each time slice. 
The dashed curves are plotted for $\rho/\rho_{\rm max}=0.01$ and 0.001.  
}
\end{figure}

\begin{figure}
\vspace*{-6mm}
\begin{center}
\leavevmode
\psfig{file=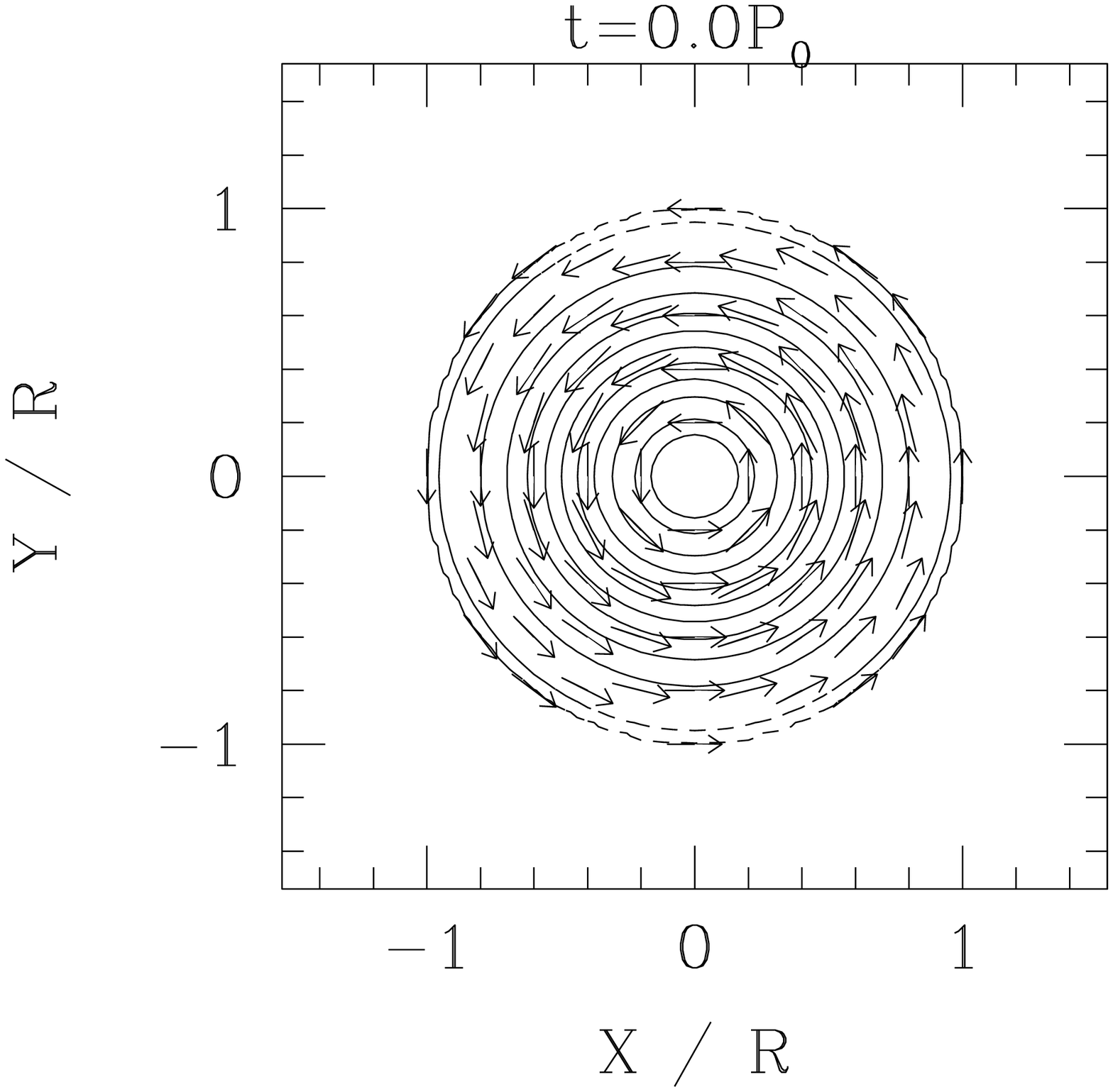,width=1.5in,angle=0}
\leavevmode
\psfig{file=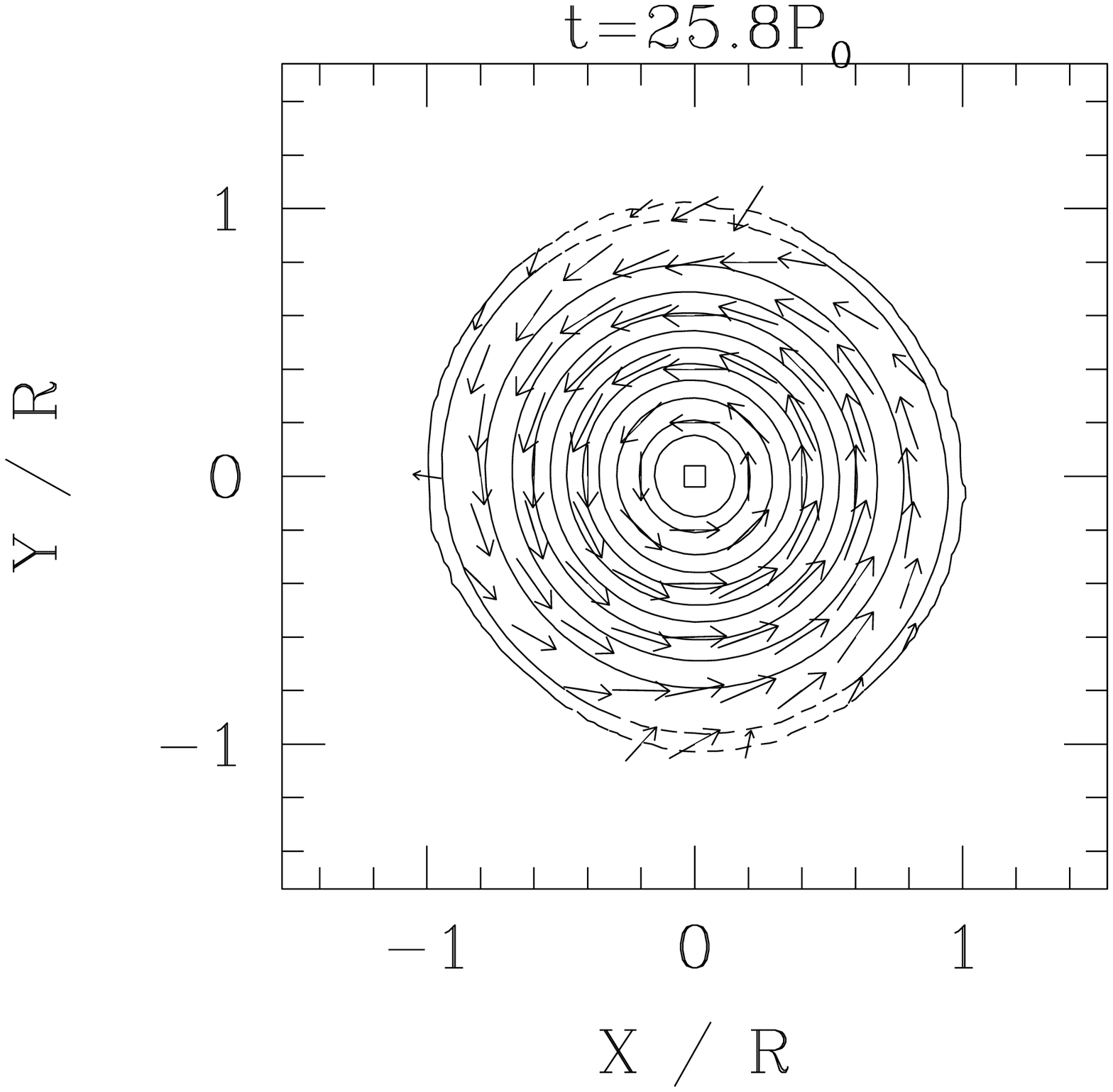,width=1.5in,angle=0}
\leavevmode
\psfig{file=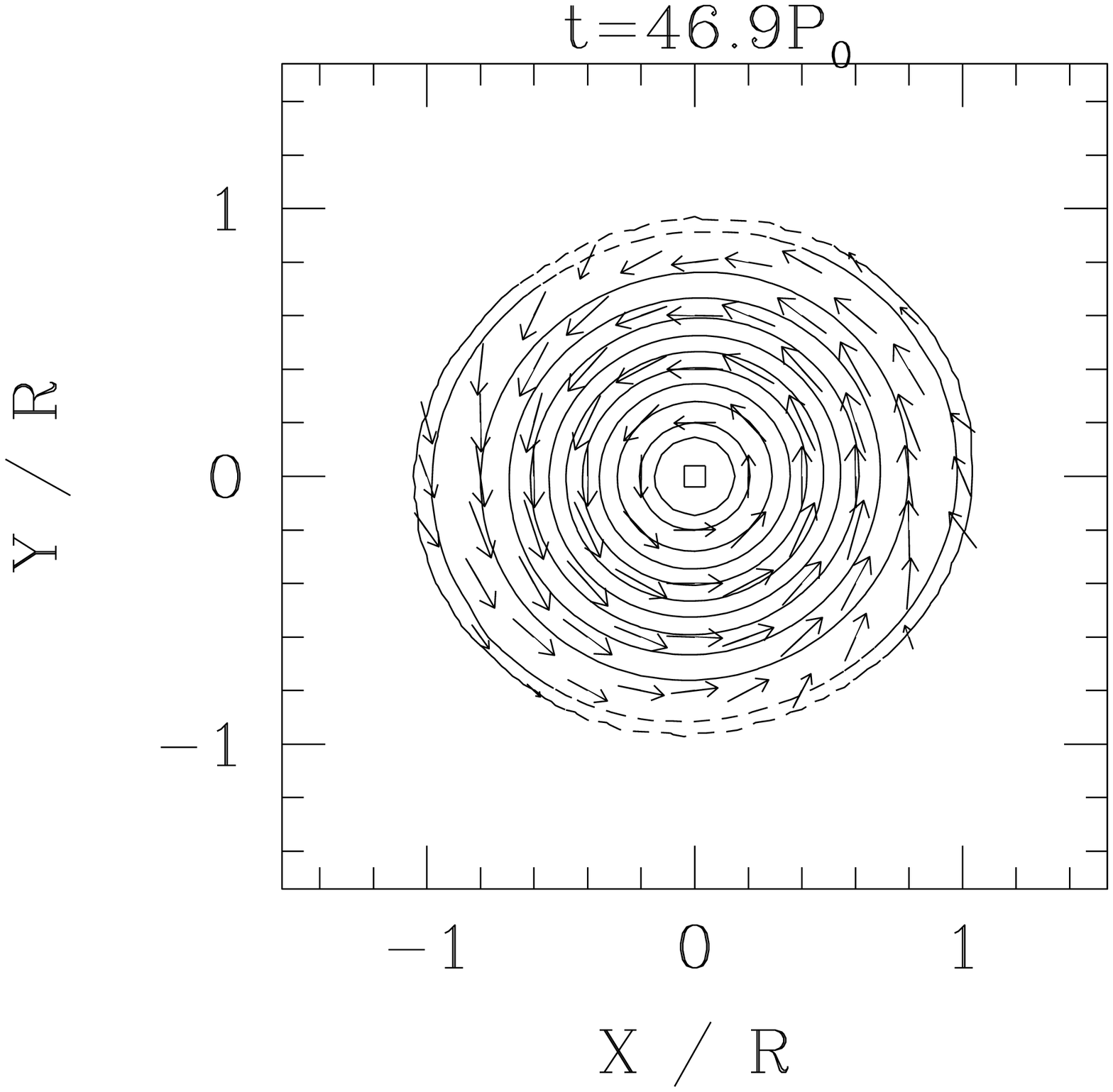,width=1.5in,angle=0}
\leavevmode
\psfig{file=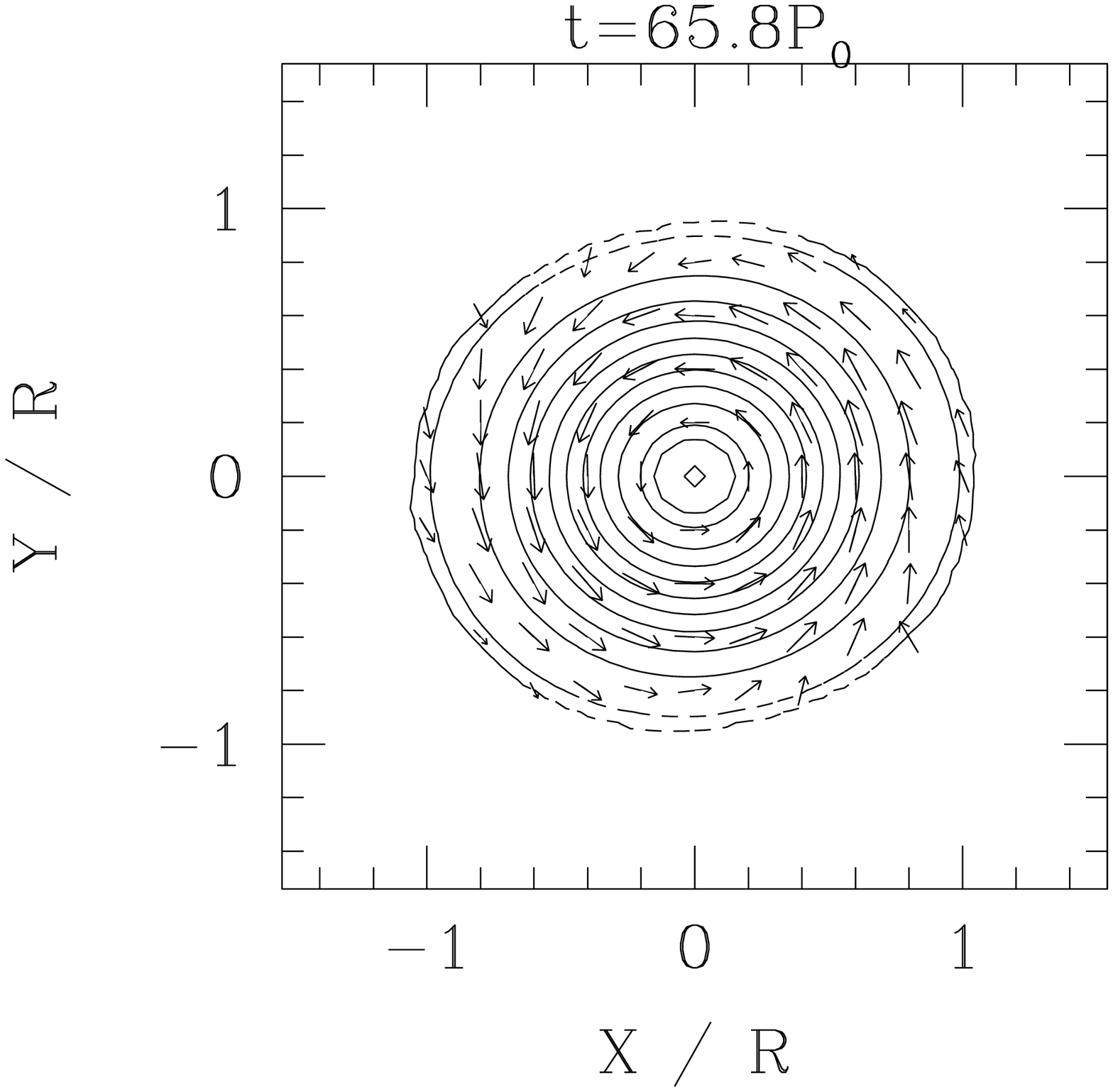,width=1.5in,angle=0}
\end{center}
\vspace*{-8mm}
\caption{The same as Figure 8, but 
for $\Gamma=5/3$, $\hat A=0.3$ and $C_a=0.8$.
}
\end{figure}

\begin{figure}
\vspace*{-6mm}
\begin{center}
\leavevmode
\psfig{file=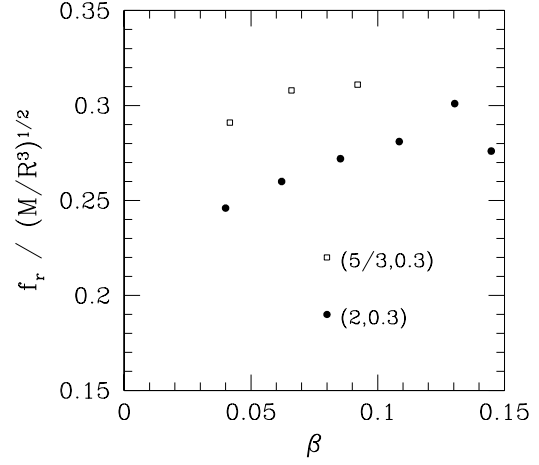,width=2.9in,angle=0}
\end{center}
\vspace*{-6mm}
\caption{$\bar f_r$ of the ellipsoids formed after onset of 
dynamical instability for the $j$-constant-like angular velocity profile. 
Filled circles and open squares denote the results for 
$(\Gamma, \hat A)=$(2, 0.3) and (5/3, 0.3). 
\label{FIG10}
}
\end{figure}

\begin{figure}
\vspace*{-6mm}
\begin{center}
\leavevmode
\psfig{file=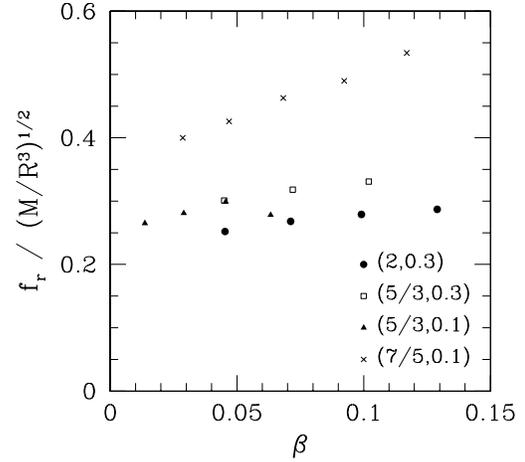,width=2.9in,angle=0}
\end{center}
\vspace*{-6mm}
\caption{The same as Figure \ref{FIG10}, but 
for the Kepler-like angular velocity profile.
Filled circles, open squares, filled triangles and crosses denote
the results for $(\Gamma, \hat A)=$(2, 0.3), (5/3, 0.3),
(5/3, 0.1) and (7/5, 0.1). 
\label{FIG11}
}
\end{figure}

In Figures \ref{FIG10} and \ref{FIG11}, we show the frequency $f_r$ of
the oscillation of the ellipsoidal star in units of
$(M/R_{\rm eq}^3)^{1/2}$ as a 
function of initial values of $\beta$ for various sets of $\Gamma$ and 
$\hat A$. The value of $f_r$ is determined by the Fourier transform of 
$\eta_{+}$ in the time domain. We note that the rotational period of 
the ellipsoid is $2 /f_r$, and that the frequency of gravitational waves is 
$f_r$.  It is interesting to note that for $\Gamma=2$ and 5/3, 
a nondimensional quantity $\bar f_r \equiv f_r (R_{\rm eq}^3/M)^{1/2}$
is in a narrow range between $0.2$ and $0.35$ 
irrespective of $\hat A$, $\beta$ and angular velocity profile. 
For $\Gamma=7/5$, $\bar f_r$ is between $0.4$ and $0.55$, which
is larger than the values for $\Gamma=2$ and 5/3.
However, it is still in a narrow range. 
The fact that the value of $\bar f_r$ depends weakly on $\beta$ suggests 
that the excited mode may be the $f$ mode. 

\subsection{Gravitational waves}

\begin{figure}
\vspace*{-6mm}
\begin{center}
\leavevmode
\psfig{file=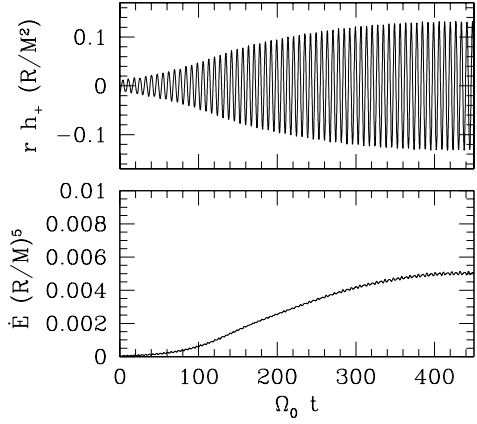,width=2.7in,angle=0}
\end{center}
\vspace*{-6mm}
\caption{Gravitational waves for $h_{+}$ in units of $M^2/R_{\rm eq}$
and the luminosity of gravitational waves
$\dot E$ in units of $(M/R_{\rm eq})^5$ as a function of $\Omega_0 t$ 
for $(\Gamma, \hat A)=(2, 0.3)$ and $C_a=0.705$ ($\beta \approx 0.071$).
\label{FIG12}
}
\end{figure}

\begin{figure}
\vspace*{-6mm}
\begin{center}
\leavevmode
\psfig{file=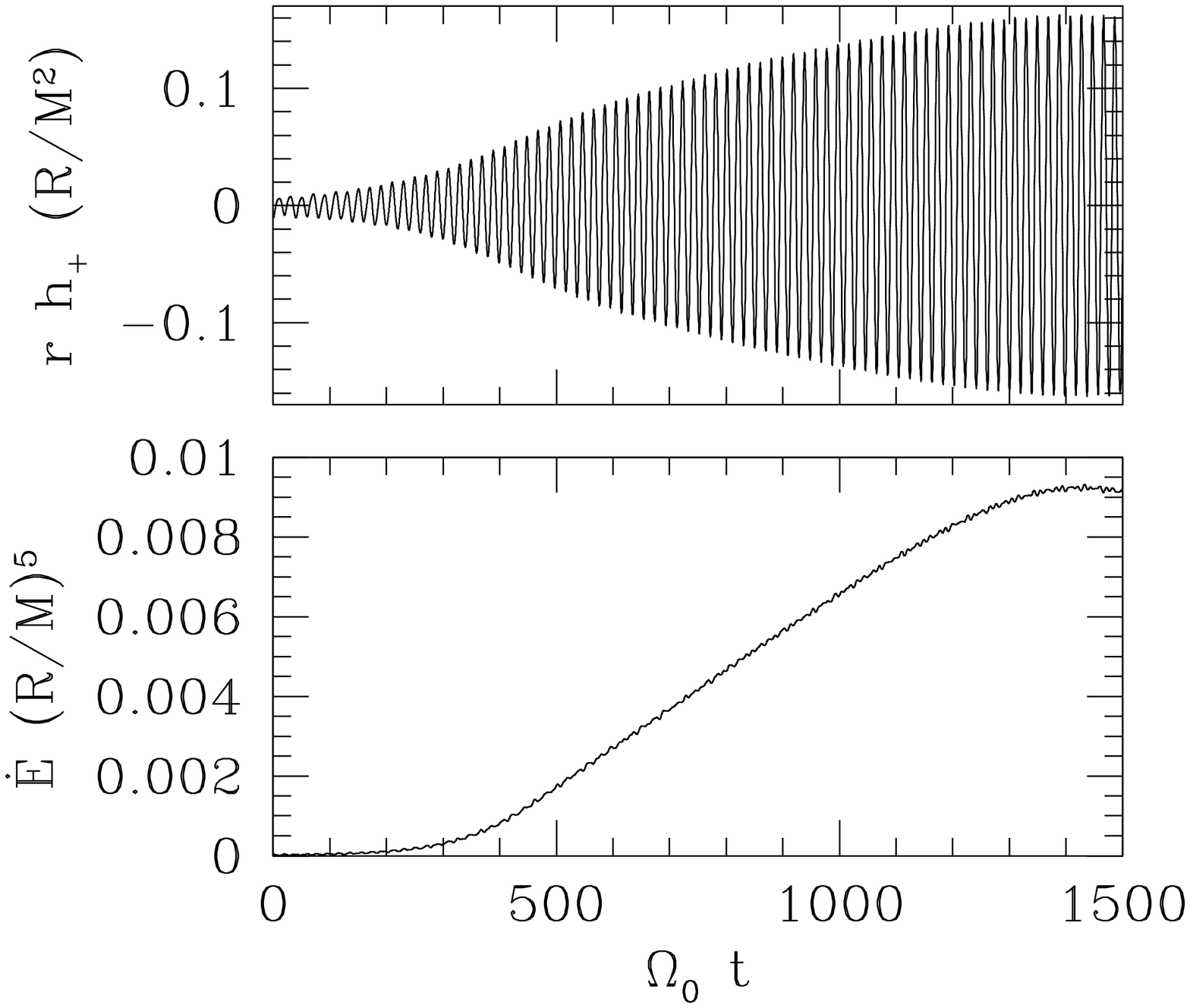,width=2.7in,angle=0}
\end{center}
\vspace*{-6mm}
\caption{The same as Figure \ref{FIG12}, but 
for $(\Gamma, \hat A)=(5/3, 0.1)$ and 
$C_a=0.705$ ($\beta \approx 0.046$). \label{FIG13}
}
\end{figure}

\begin{figure}
\vspace*{-6mm}
\begin{center}
\leavevmode
\psfig{file=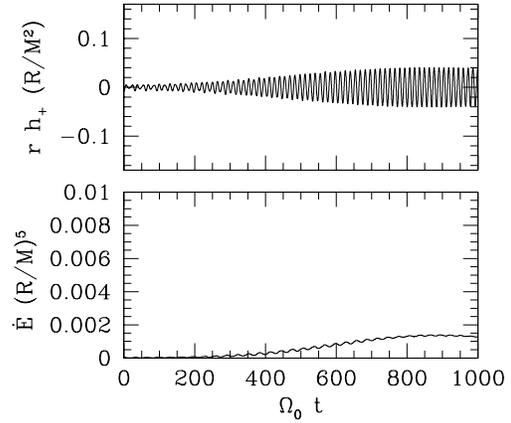,width=2.7in,angle=0}
\end{center}
\vspace*{-6mm}
\caption{The same as Figure \ref{FIG12}, but 
for $(\Gamma, \hat A)=(7/5, 0.1)$ and 
$C_a=0.705$ ($\beta \approx 0.047$). \label{FIG14}
}
\end{figure}

As discussed in our previous paper \cite{SKE}, the dynamically unstable 
rotating stars, which are deformed to nonaxisymmetric ellipsoidal objects, 
are likely to be sources of quasi-periodic gravitational waves. In Figures
\ref{FIG12}--\ref{FIG14},  
we show the gravitational waveforms of the $+$ mode along the $z$-axis 
($h_{+}$) and the luminosity ($\dot E$) as a function of time for 
$(\Gamma, \hat A)=(2, 0.3)$, (5/3, 0.1) and (7/5, 0.1). For all the models 
picked up here,  $C_a=0.705$. The value of $\beta$ is $\approx 0.071$,
0.046, and 0.047, respectively. 
We note that the waveforms for the $\times$ mode 
are essentially the same as those for the $+$ mode except for the
phase difference by $\pi/4$. Here, we calculate gravitational waves in the
quadrupole formula \cite{MTW} and define the waveforms by 
\beqn
h_+ \equiv {\ddot I_{xx} - \ddot I_{yy} \over r},~
h_{\times} \equiv {2\ddot I_{xy} \over r},
\eeqn
and the luminosity by 
\beqn
\dot E \equiv {1 \over 5} \sum_{i,j} \bI_{ij}^{(3)}\bI_{ij}^{(3)}, 
\eeqn
where
\beqn
&& \bI_{ij}=I_{ij}-{\delta_{ij} \over 3} \sum_{k} I_{kk}, \nonumber \\
&& \ddot I_{ij}={d^2 I_{ij} \over dt^2}, ~~~~~
\bI_{ij}^{(3)}={d^3 \bI_{ij} \over dt^3},
\eeqn
and $r$ is the distance from a source to a detector. $h_+$ and $h_{\times}$
are the waveforms observed along the $z$ axis. 

In the early phase before the growth of the nonaxisymmetric perturbation 
saturates, the amplitude of gravitational waves increases
gradually together with 
the magnitude of $\eta$. Then, the growth of the amplitude saturates and 
subsequently, quasi-periodic gravitational waves are emitted for 
a much longer time than the rotational period. The amplitude of 
quasi-periodic gravitational waves depends on $\Gamma$ and $\hat A$, as in 
the case of the saturated value of $\eta$. For a given set of $\hat A$ and 
$C_a$ (or $\beta$),
it is smaller for smaller values of $\Gamma$. This is
because the star with smaller values of 
$\Gamma$ is more centrally condensed and, hence,  
the magnitudes of the quadrupole moments are smaller for
the identical mass and radius.

We found that 
for the parameters we studied 
($\beta \alt 0.1$, $\hat A \leq 0.6$ and angular velocity profiles
(1) and (2)), 
the typical values of the amplitude and luminosity of gravitational
waves for $\Gamma=2$ are $r h_{+,\times} \sim 0.1$--$0.2(M^2/R_{\rm eq})$ 
and $\dot E \sim 0.005$--$0.01 (M/R_{\rm eq})^5$.
For $\Gamma=5/3$, the magnitudes of them
are only slightly smaller than those for $\Gamma=2$. 
However, for $\Gamma=7/5$, they are smaller by a factor of several as
$rh_{+,\times} \sim 0.05(M^2/R_{\rm eq})$
and $\dot E \sim 0.001 (M/R_{\rm eq})^5$. 

Using the results shown in Figures \ref{FIG12}--\ref{FIG14}, 
we can estimate the expected effective amplitude of gravitational waves
from the nonaxisymmetric outcomes formed after the dynamical instability
saturates. Here, we pay particular attention to proto-neutron stars which
are likely formed soon after the supernovae of mass $\sim 1.4M_{\odot}$ 
and radius several 10 km. For the luminosity 
$\dot E = 0.001\epsilon(M/R_{\rm eq})^5$
where $\epsilon$ is a parameter of magnitude 1--10, the emission timescale 
of gravitational waves can be estimated as
\beqn
\tau \sim {T \over \dot E} &=& 100 \alpha_0
\epsilon^{-1} \biggl({\beta \over 0.1}\biggr)
\biggl({R_{\rm eq} \over M}\biggr)^4 M \nonumber \\
&=& 6.1~{\rm sec} \ \ \alpha_0
\epsilon^{-1} \biggl({\beta \over 0.1}\biggr)
\biggl({R_{\rm eq} \over 30~{\rm km}}\biggr)^4 
\biggl({M \over 1.4M_{\odot}}\biggr)^{-3},  
\eeqn
where we set $T = \alpha_0 \beta M^2/R_{\rm eq}$,
and $\alpha_0$ is a constant 
which depends on the value of $\Gamma$ but very weakly on $\hat A$; 
for $\beta \alt 0.1$, $\alpha_0 \sim 0.8$, 0.9, and 1.2
for $\Gamma=2$, 5/3, and 7/5, respectively, within $\sim 10\%$ error.

The characteristic frequency of gravitational waves is denoted as 
\beqn
f = f_r
\approx 790~{\rm Hz}~ \biggl({\bar f_r \over 0.3}\biggr)
\biggl({R_{\rm eq} \over 30~{\rm km}}\biggr)^{-3/2}
\biggl({M \over 1.4M_{\odot}}\biggr)^{1/2}. 
\eeqn
Assuming that the nonaxisymmetric perturbation would not be dissipated
by viscosities or magnetic fields on the emission timescale of gravitational 
waves \cite{BSS}, the accumulated cycles of gravitational wave-train $N$
are estimated as 
\beqn
N \equiv f\tau &=&4.8\times10^3 \alpha_0
\biggl({\epsilon \over 5}\biggr)^{-1}
\biggl({\bar f_r \over 0.3}\biggr) \nonumber \\
&\times&
\biggl({\beta \over 0.1}\biggr) 
\biggl({R_{\rm eq} \over 30~{\rm km}}\biggr)^{5/2}
\biggl({M \over 1.4M_{\odot}}\biggr)^{-5/2}. 
\eeqn
The effective amplitude of gravitational waves is defined by 
$h_{\rm eff} \equiv N^{1/2}h$ where $h$ denotes the characteristic amplitude 
of periodic gravitational waves. Using this relation, we find 
\beqn
h_{\rm eff}&\approx&
3.2 \times 10^{-22} \alpha_0^{1/2}
\biggl({\bar h \over 0.1}\biggr)
\biggl({\epsilon \over 5}\biggr)^{-1/2}
\biggl({\beta \over 0.1}\biggr)^{1/2}
\biggl({\bar f_r \over 0.3}\biggr)^{1/2}\nonumber \\
&\times &
\biggl({R_{\rm eq} \over 30~{\rm km}}\biggr)^{1/4} 
\biggl({M \over 1.4M_{\odot}}\biggr)^{3/4}
\biggl({100~{\rm Mpc} \over r}\biggr) 
\eeqn
\cite{Kip,LS,LL1,LL2} where $\bar h\equiv h r R_{\rm eq}/M^2$.
Since $\bar f_r$, $\bar h$, $\epsilon$ and $\beta$ depend on the values of 
$\Gamma$, $\hat A$, and $C_a$, $h_{\rm eff}$ can vary by a factor of
$\sim 3$.  
However, for all the rotating stars that we studied in this paper,
$h_{\rm eff}$ is always larger than $10^{-22}$ at a distance 
$r \sim 100$ Mpc with $R_{\rm eq} \sim 30$ km and $M \sim 1.4M_{\odot}$. 
Furthermore, the frequency of gravitational waves is about 1 kHz
for $R_{\rm eq} \sim 30$ km and $M \approx 1.4M_{\odot}$. 
Thus, gravitational waves from proto-neutron stars of
a high degree of differential rotation, of mass
$\sim 1.4 M_{\odot}$, and of radius $\agt 30$ km at a 
distance of $\sim 100$ Mpc are likely to be sources for 
laser interferometric detectors such as LIGO \cite{Kip2},
if the other dissipation processes are negligible. 

\section{Summary and discussion}

We have studied the dynamical bar-mode instability of differentially rotating 
stars of polytropic equations of state. We chose three polytropic
indices and two angular velocity profiles in this study.
We found that rotating stars of a high degree of differential 
rotation are dynamically unstable against the nonaxisymmetric bar-mode
deformation
even with $\beta \ll 0.27$, irrespective of the polytropic indices and 
angular velocity profile. The criterion of the value of $\beta$ for onset of 
the instability depends on the rotational profile and the equations of state, 
but the dependence is very weak if the degree of differential 
rotation is high enough as $\hat A \sim 0.1$. 

We estimated the effective amplitude of gravitational waves from 
nonaxisymmetric objects formed after onset of the dynamical instability. 
For typical proto-neutron stars of mass $\sim 1.4 M_{\odot}$ and radius 
several $10$ km, the effective amplitude of gravitational waves 
at a distance of $\sim 100$ Mpc is larger than $10^{-22}$, and 
the frequency $\sim$ 1 kHz. Therefore, 
the gravitational waves can be sources for laser interferometric 
detectors such as advanced LIGO (e.g., Thorne 1995). 

As we mentioned above, this conclusion is drawn under the assumption that 
dissipation of nonaxisymmetric perturbations by viscosity and magnetic fields 
is negligible. The dissipation timescale due to molecular viscosities and 
magnetic braking is likely to be longer than 10 sec \cite{BSS}. Thus, these 
effects can be safely neglected. However, turbulent magnetic 
viscosity \cite{BH} may be relevant for redistribution of the angular 
momentum profile. This implies that a differentially rotating star might 
be enforced to a rigidly rotating state on a dynamical timescale
and, hence, the 
nonaxisymmetric structure might disappear. Many theoretical works have 
clarified that the magnetic viscous effect can redistribute the
angular momentum distribution of accretion disks around central
objects on a dynamical timescale \cite{BH}. However, to our knowledge, there 
is no work on the magnetic field effect to self-gravitating rotating stars.
The magnetic field would redistribute the angular momentum of the 
differentially rotating stars on a dynamical timescale, but it is not clear 
if it is strong enough to enforce the rotating stars to an
axisymmetric state in a dynamical timescale.  To clarify this problem, 
it is necessary to carry out MHD simulations for rotating stars. Such work 
should be done in the future.

\section*{Acknowledgments}

We thank L. Lindblom and I. Hachisu for discussion and comments. 
Numerical simulations were performed on FACOM VPP5000 in the 
data processing center of National Astronomical Observatory of Japan. 
This work was in part supported by a Japanese 
Monbu-Kagaku-Sho Grant (Nos. 12640255, 13740143 and 14047207). 
SK is supported by JSPS Research Fellowship for Young Scientists.

\label{lastpage}

\end{document}